\documentclass[nofootinbib,aps,prd,groupedaddress,preprint]{revtex4-1}
\usepackage{amssymb,amsmath,amsbsy}
\usepackage{mathrsfs,verbatim,enumerate}
\usepackage{slashed}
\usepackage[dvips]{graphics}
\usepackage{epsfig}
\usepackage{subfigure}

\setcitestyle{numbers}

\def\beq{\begin{equation}}
\def\eeq{\end{equation}}
\def\beqa{\begin{eqnarray}}
\def\eeqa{\end{eqnarray}}
\def\ifmath#1{\relax\ifmmode #1\else $#1$\fi}
\def\lsup#1{^{\lower 4pt\hbox{$\scriptstyle#1$}}}
\def\llsup#1{^{\lower 2pt\hbox{$\scriptstyle#1$}}}

\def\lsim{\mathrel{\raise.3ex\hbox{$<$\kern-.75em\lower1ex\hbox{$\sim$}}}}
\def\gsim{\mathrel{\raise.3ex\hbox{$>$\kern-.75em\low
er1ex\hbox{$\sim$}}}}

\def\fig#1{Fig.~\ref{#1}}

\newcommand{\MET}{\mbox{$E\kern-0.50em\raise0.10ex\hbox{/}_{T}$}}
\newcommand{\METzero}{\mbox{$E\kern-0.50em\raise0.10ex\hbox{/}_{T}^{0}$}}
\newcommand{\vecMET}{\mbox{$\vec{E}\kern-0.50em\raise0.10ex\hbox{/}_{T}$}}
\newcommand{\METSPEC}{\mbox{$E\kern-0.50em\raise0.10ex\hbox{/}_{Tspec}$}}

\def\slantfrac#1#2{\kern.1em^{#1}\kern-.3em/\kern-.1em_{#2}}

\newcommand{\gammaslash}[1]{#1\kern-0.45em/}
\newcommand{\Wslash}{\mbox{$W\kern-0.80em/\ $}}
\newcommand{\Wslashdag}{\mbox{$W\kern-0.80em\raise0.10ex\hbox{/ }^{\dagger}$}}
\newcommand{\Dslash}{\mbox{$D\kern-0.70em/$}}
\newcommand{\Bslash}{\mbox{$B\kern-0.70em/\ $}}
\newcommand{\pslash}{\mbox{$p\kern-0.45em/\ $}}
\newcommand{\Partslash}{\mbox{$\partial\kern-0.55em/$}}

\begin{document}
\pagestyle{empty}
\title{\Large\bf Prospects for Measuring the Mass of Heavy, Long-lived Neutral Particles that Decay to Photons}

\author{Ziqing Hong}
\email{zqhong@tamu.edu}
\affiliation{Mitchell Institute for
Fundamental Physics and Astronomy, Texas A\&M University, College Station TX 77843-4242}
\author{David Toback}
\email{toback@tamu.edu}
\affiliation{Mitchell Institute for
Fundamental Physics and Astronomy, Texas A\&M University, College Station TX 77843-4242}
\date{\today}







\begin{abstract}
We describe a novel way to measure the mass of heavy, long-lived neutral particles that decay to photons using collider experiments. We focus on a Light Neutralino and Gravitino model in a Gauge Mediated Supersymmetry Breaking scenario where the neutralino has a long-lifetime ($\mathcal{O}$(ns)) as it is not excluded by current experiments. To illustrate our method and give sensitivity estimates we use recent CDF results and a production mechanism where sparticles are produced via $\phi_{i} \rightarrow \widetilde{\chi}^{1}_{0} \widetilde{\chi}^{1}_{0} \rightarrow (\gamma \widetilde{G})(\gamma \widetilde{G})$ in which $\phi_{i}$ indicates a neutral scalar boson, $\widetilde{\chi}^{1}_{0}$ is the lightest neutralino and $\widetilde{G}$ is the gravitino, as a full set of background shapes and rates are available. Events can be observed in the exclusive photon plus Missing $E_{T}$ final state where one photon arrives at the detector with a delayed time of arrival. Surprisingly, a simple measurement of the slope of the delayed-time distribution with the full CDF dataset is largely insensitive to all but the $\widetilde{\chi}^{1}_{0}$ mass and allows for the possibility of determining its mass to approximately 25\% of itself.
\end{abstract}


 \maketitle

\newpage
\setcounter{page}{1}
\setcounter{footnote}{0}
\pagestyle{plain}

\section{Introduction}

While the standard model of particle physics~\cite{Glashow1961579,*Weinberg:1967tq,*sm_salam} has been enormously successful, there are many reasons to believe it is not a final theory of nature. For example, the hierarchy problem~\cite{PhysRevD.19.1277,*PhysRevD.20.2619,*PhysRevD.13.974,*PhysRevD.14.1667,*hooft1980recent} and the possibility of astronomical observations that can be interpreted in cosmological models as being evidence for a 4th relativistic species~\cite{0004-637X-730-2-119,0067-0049-192-2-18} are suggestive of new physics. Many models have been proposed to solve these problems, and others, by embedding of the Higgs potential into a Minimal Supersymmetric Standard Model~(MSSM)~\cite{Dawson:1997tz}. A full set of searches is underway to discover the new particles predicted by these models around the world so far without success~\cite{pdg} indicating the need to follow up on less common scenarios. An important such possibility is that resonant production of neutral scalars can produce pairs of heavy, long-lived neutral particles that decay into a photon and a new weakly-interacting particle~\cite{Mason:2011zv}. In this paper we describe a novel technique for measuring the mass of the heavy long-lived neutral particle if it is discovered in a collider experiment. For concreteness, we will focus on a simple model that is both not ruled out by experiment, has some appealing theoretical aspects, and allows full sensitivity study as we have a full set of recent search results from CDF data at the Tevatron. Note that our results ought to be scalable to other collider experiments, like the LHC, or for other final state topologies, such as associated production of heavy, long-lived neutral particles that decay to photons. To keep our discussion quantitative, estimates from other experiments are outside the scope of this paper, but can be done when background rates and signal shapes become available.

\section{Baseline Model}

For pedagogical reasons we begin by outlining a baseline model that illustrates the final state we are considering in a collider experiment, and after that is done we describe a novel way to measure the mass of the massive, neutral long-lived particle if such a particle were to be discovered. We note at the outset while that many of the individual techniques here are not new, we found that despite how complicated the possibilities were, using only simple measurements of the data, a fairly robust mass measurement is possible, and this measurement is insensitive to many of the other parameters of the model. In the interests of concreteness, we have focused on making quantitative estimates of the sensitivity using only experimental results for reliable information. 

With this in mind, we focus on a Gauge Mediated Supersymmetry Breaking~(GMSB) scenario~\cite{Dine1981575,Dimopoulos1981353,Dine1982227,Nappi1982175,AlvarezGaumé198296,Dimopoulos1983479} as it is one of many scenarios that has the theoretical advantages of being able to solve the above problems (as well as others discussed in the literature), as well as because it is easy to work with and makes concrete predictions for judicious choices of the parameters. In this model, the lightest neutralino, $\widetilde{\chi}^{1}_{0}$, is often the Next-to-Lightest Supersymmetric Particle (NLSP), and the Gravitino, $\widetilde{G}$, is the Lightest Supersymmetric Particle (LSP), with a mass less than a ${\rm KeV/c^2}$~\cite{PhysRevD.48.1277,PhysRevD.51.1362,PhysRevD.53.2658,PTPS.177.143,PhysRevD.79.035002,1126-6708-2009-03-016,1126-6708-2009-03-072,Rajaraman2009367,springerlink:10.1007/JHEP06(2010)047,springerlink:10.1007/JHEP05(2010)096}; the light $\widetilde{G}$ can serve as an additional relativistic species of particles in the early universe. The production and decay of $\widetilde{\chi}^{1}_{0}\rightarrow\gamma+\widetilde{G}$ can lead to interesting $\gamma + \slashed{E}_{T}$ final states if sparticles are produced at colliders~\cite{PhysRevLett.76.3494,PhysRevLett.76.3498,PhysRevD.54.5395,PhysRevD.54.3283,Dimopoulos199739,PhysRevLett.77.5168,PhysRevD.55.4463,PhysRevD.56.1761,
PhysRevD.68.014019,PhysRevD.69.035003,Shirai2009351,JHEP05,springerlink:10.1007/JHEP10(2010)067} and the $\widetilde{\chi}^{1}_{0}$ mass is lower than the $Z^0$ mass ($m_{\widetilde{\chi}^{1}_{0}} < m_{Z^0}$).

While there is no evidence for GMSB production and decay in experiments~\cite{springerlink:10.1007/s10052-002-1005-z,*2652040520070420,*Pasztor:2005es,*springerlink:10.1140/epjc/s2004-02051-8,PhysRevLett.104.011801,
*PhysRevLett.105.221802,PhysRevLett.99.121801,*PhysRevD.78.032015,PhysRevLett.106.211802,*PhysRevLett.106.121803,*Chatrchyan:2012ir,*Chatrchyan:2012jwg,*PhysRevD.88.012001}, there are some interesting scenarios that have not been ruled out. The current experimental searches for GMSB at collider experiments consider both short lifetimes ($\tau_{\widetilde{\chi}^{1}_{0}} < 1$ns) and longer lifetimes (2 ns $< \tau_{\widetilde{\chi}^{1}_{0}} < $50 ns), but these searches usually assume a minimal GMSB model and use the SPS-8 relations~\cite{SPS-8} because they yield large cross section values for direct production of gaugino pairs and/or production of squarks and gluinos. When the SPS-8 relations are released, and more general GMSB model scenarios are considered, an important variation is that the $\widetilde{\chi}^{1}_{0}$ and the $\widetilde{G}$ can be the only kinematically accessible supersymmetric particles at colliders~\cite{higgs2009}. In this case, known as the Light Neutralino and Gravitino (LNG) scenario, current sparticle mass bounds are evaded. A study of the phenomenology of this scenario at colliders~\cite{Mason:2011zv} indicates that if the $m_{\widetilde{\chi}^{1}_{0}}$ is less than half the masses of any new neutral scalar bosons that exist in nature, $\phi_i$, and if the couplings are favorable, sparticle production can be dominated by resonant production of $\phi_i \rightarrow \widetilde{\chi}^{1}_{0} \widetilde{\chi}^{1}_{0}$. This could readily produce anomalous production of events in the $\gamma + \slashed{E}_{T}$ final state. Note that in Ref.~\cite{Mason:2011zv} associated vector boson production diagrams are not considered as they have smaller production cross sections (and background estimates have not been done).


The number of neutral scalar bosons in nature is not well constrained by experiments. There is clear evidence for at least one ``Higgs Like" boson with a mass around $125~{\rm GeV/c^2}$~\protect{\cite{PhysRevLett.109.071804,higgs_CMS,higgs_ATLAS}} with some of the properties of the SM Higgs boson~\cite{MingshuiChenfortheCMS:2013fba,*Consonni:2013zsa}, but most versions of Supersymmetry (SUSY) require at least a two-Higgs doublet~\cite{PhysRevD.39.844}. While searches for other MSSM Higgs bosons have found no evidence so far~\cite{Lai:2013yaa,*Schael:2006cr,*Benjamin:2010xb,*Abazov:2011up,*Aad:2012cfr,*Chatrchyan:2011nx,*Aad:2011rv}, the discrepancy of the scalar boson mass measured in $ZZ^{(*)} \rightarrow 4l$ channel and in $\gamma \gamma$ channel at ATLAS~\cite{higgs_ATLAS} could be a sign of the existence of multiple Higgs bosons with masses around $125~{\rm GeV/c^2}$ (although it is not clear if this hint should be taken seriously). Either way, it is not unreasonable to consider the case of the production of $\widetilde{\chi}^{1}_{0}$ pairs from a neutral scalar boson, and contemplate large values of cross section times branching fraction when considering one or more species of neutral scalar bosons that contribute to the production diagrams. This could occur if the couplings are favorable or if multiple scalers exist in nature. For our baseline model we will simplify and refer to single scalar production with $\phi \rightarrow \widetilde{\chi}^{1}_{0}\widetilde{\chi}^{1}_{0}$.

While the lifetime of a heavy long-lived particle is typically unconstrained, as is the case of a $\widetilde{\chi}^{1}_{0}$ in GMSB scenarios, we focus on the long-lived $\widetilde{\chi}^{1}_{0}$ scenario for a number of reasons. If there were $\phi \rightarrow \widetilde{\chi}^{1}_{0} \widetilde{\chi}^{1}_{0}\rightarrow (\gamma \widetilde{G})(\gamma \widetilde{G})$ production, this final state should yield production of two photons and $\slashed{E}_{T}$ in the final state, although the number of photons identified in a detector is dependent on the lifetime of the neutralino~\cite{SearchProspect}. Previous studies of this type of production and decay~\cite{SearchProspect} indicate that the most likely way to be sensitive to this type of final state is the exclusive final state because of the fake $\slashed{E}_{T}$ backgrounds that can arise.

There are no direct searches published in the LNG scenario. LEP wouldn't have large direct production capabilities, and small number of final state particles and low $E_{T}$ kinematics, coupled with the large number of interactions per crossing at the LHC may make it less sensitive to these models, which might explain why there are no results out in this final state yet. But CDF has started considering them. Specifically, a CDF result in exclusive $\gamma\gamma + \slashed{E}_{T}$, which is expected to be sensitive to $\tau_{\widetilde{\chi}^{1}_{0}} < 2$~ns, didn't show an excess~\cite{PhysRevD.82.052005}. There are currently no limits in the case of a long-lived neutralino, but there is a recently released result from CDF in the exclusive $\gamma + \slashed{E}_{T}$ final state using photons that arrive late compared to expectations, so-called delayed photons~\cite{cdf10788.2,*JonThesis,*AdamThesis}. We note that most of the data in the signal region appears to be above the backgrounds in a suggestive way, albeit not in one that is statistically significant. With a clear indication that such production can be searched for, and a clear way to estimate background rates, we focus the rest of our study on this scenario as we can estimate signal and backgrounds for a full sensitivity study. We move to specify the parameter space we will consider within our baseline model so that we can determine what, in principle, could be measured in the case of an observed excess of events. As we will see, a novel method for determining the mass of the $\widetilde{\chi}^{1}_{0}$ will be possible.

\section{Parameter space with the Baseline Model}

To illustrate the sensitivity of measuring the mass of a heavy, long-lived neutral particle, we consider the region of parameter space in the baseline model that is both not excluded by current experiments and would allow for a measurement. As there already exist sensitivity estimates for discovery of $\widetilde{\chi}^{1}_{0}$ pair production from a SM Higgs boson in the exclusive $\gamma + \slashed{E}_{T}$ final state at the Tevatron~\cite{Mason:2011zv}, we follow that model. However, since that report pre-dates the discovery of a Higgs-like boson, we begin by extending our baseline model to include the possibility of neutralino pairs being produced from any number of neutral scalars with similar masses. While the production cross section for each boson and its branching fraction to $\widetilde{\chi}^{1}_{0}$ pairs is critically dependent on the masses and couplings of each, as will be seen, our method is not terribly sensitive to this assumption, although it will affect the statistical uncertainties of any measurement.

There are multiple parameters in the LNG scenario of GMSB, each of which affect the masses of the neutral scalar bosons, their couplings to SM and SUSY particles as well as the mass and the lifetime of $\widetilde{\chi}^{1}_{0}$. These in turn affect the cross sections, the branching fractions to $\widetilde{\chi}^{1}_{0}$ pairs, as well as the kinematics and other observables of the final state. For phenomenological purposes, we can consider the production and decay of any set of series of neutral scalar bosons as a single scalar production, with a fixed mass if the masses are similar. 
Thus, from here on we will simply discuss our model as if it were single production and decay of a SM Higgs boson, denoted as $\mathrm{\varphi}$.

With these simplifying assumptions, the number of effective observable free parameters in the baseline scenario is narrowed down to four: the effective production cross section of neutralino pairs ($\sigma_{\widetilde{\chi}^{1}_{0} \widetilde{\chi}^{1}_{0}}$), the effective scalar mass ($m_{\mathrm{\varphi}}$), the neutralino mass ($m_{\widetilde{\chi}^{1}_{0}}$), and the neutralino lifetime ($\tau_{\widetilde{\chi}^{1}_{0}}$). The rate of $\sigma_{\widetilde{\chi}^{1}_{0} \widetilde{\chi}^{1}_{0}}$ only has impact on the rate of production while other three parameters mostly affect the acceptance and the observables in the final state.

To make concrete sensitivity estimates, we choose an effective cross section to use. We note that the lightest Higgs boson, if it has SM-like couplings, has a predicted production cross section of about a picobarn at the Tevatron~\cite{1126-6708-2009-04-003,*deFlorian2009291} if we assume a mass of $125~{\rm GeV/c^2}$. On the flip side, interpreting the observed boson as the Higgs boson yields limits on its branching ratio to invisible particles to be 0.37 at 95\% C.L.~\cite{JHEP12(2012)045} which can be used to constrain the production cross section of $\widetilde{\chi}^{1}_{0}$ pairs. If we assume that other neutral scalars exist, and each scalar has a non-dominant branching fraction decaying to $\widetilde{\chi}^{1}_{0}$ pairs, then it is reasonable to estimate the prospects if $\Sigma\sigma(p\bar{p} \rightarrow \phi_i \rightarrow \widetilde{\chi}^{1}_{0} \widetilde{\chi}^{1}_{0})$ $\sim$0.5~pb where we have summed over all production and decay fractions. Any result here can then be easily scaled by changing this assumption.

We next narrow the focus to the region of parameter space that is not excluded and could produce an observable result and measurement in the data. Since recent results show evidence of a $125~{\rm GeV/c^2}$ Higgs-like particle, we focus on the region $120~{\rm GeV/c^2} \leq m_{\mathrm{\varphi}} \leq 130~{\rm GeV/c^2}$. We focus on neutralinos with less than half of this mass ($m_{\widetilde{\chi}^{1}_{0}} \leq 60~{\rm GeV/c^2}$) to keep the production cross section of $\widetilde{\chi}^{1}_{0}$ pairs high. The combination of the null result in the exclusive $\gamma\gamma+\slashed{E}_{T}$ and the hint in $\gamma_{Delayed}+\slashed{E}_{T}$ both push us to focus on $\tau_{\widetilde{\chi}^{1}_{0}} \geq$ 2ns. For reasons described in Ref.~\cite{SearchProspect}, we will not have sensitivity for light $\widetilde{\chi}^{1}_{0}$ ($m_{\widetilde{\chi}^{1}_{0}} < 30~{\rm GeV/c^2}$) or neutralino lifetimes above approximately 30~ns. This occurs because if the $m_{\widetilde{\chi}^{1}_{0}}$ is too small, or if $\tau_{\widetilde{\chi}^{1}_{0}}$ goes too large, the $\widetilde{\chi}^{1}_{0}$ is more likely to leave the detector before it decays~(invisible final state), or if it has a large boost which pushes the photon arrival time to become indistinguishable from being promptly produced.

Significant numbers of delayed photons can be observed if the neutralino lifetime is in the range of 2~ns $< \tau_{\widetilde{\chi}^{1}_{0}} < $30~ns, as is favored in some models when the SUSY breaking scale is low~\cite{PhysRevLett.76.3494}. In this case usually one neutralino will leave the detector, but the other can travel a significant distance before decaying, allowing the photon to arrive at the detector with a significant delay time relative to expectations. These photons are known as $\gamma_{Delayed}$. The final state would then be a single $\gamma_{Delayed} + \slashed{E}_{T}$, and fake backgrounds from $W\gamma \rightarrow \ell \nu \gamma \rightarrow \gamma + \ell_{\rm lost} + \slashed{E}_{T}$ and $\gamma+{\rm jet} \rightarrow \gamma +{\slashed{E}_{T}}_{\rm fake}$ make it beneficial to search in the final state of exclusive single $\gamma_{Delayed} + \slashed{E}_{T}$~\cite{Acosta:2002eq}. These delayed photons can be separated from SM and non-collision backgrounds using known techniques with the EMTiming system at CDF~\cite{Goncharov2006543,PhysRevLett.99.121801,*PhysRevD.78.032015,SearchProspect}. Specifically, a measure of the amount the photon is delayed is given by $t_{corr}~\equiv~t_f~-~t_i~-~\frac{|\vec{x}_f-\vec{x}_i|}{c}$, where $t_f$ and $\vec{x}_f$ are the time of arrival and position the photon hits the calorimeter respectively, and $t_0$ and $\vec{x}_0$ are the collision time and position respectively.


Before continuing, we comment quickly more on the $t_{corr}$ distribution of the recent CDF result in the exclusive $\gamma_{Delayed}+\slashed{E}_{T}$ result with  6.3 $~{\rm fb^{-1}}$ data~\cite{cdf10788.2,*JonThesis,*AdamThesis}. They report 322 observed events in the timing window 2~ns $\leq t_{corr} \leq$ 7~ns (signal region) on a background of $286 \pm 24$ for a significance of $1.2\sigma$. While the quoted result is not statistically significant, and one should not take it too seriously, we do note that the shape of the $t_{corr}$ distribution (after background subtraction) looks very much like the shape we expect from the LNG scenario of GMSB, a falling exponential on top of the SM and cosmic ray backgrounds.

With a specific region in our baseline well specified, we come to the question of what we can measure if $p\bar{p} \rightarrow \phi_{i} \rightarrow \widetilde{\chi}^{1}_{0} \widetilde{\chi}^{1}_{0} \rightarrow \gamma_{Delayed}+\slashed{E}_{T}$ were discovered. In particular, we estimate both the expected number of events for various model parameter choices as well as lay out our method of analyzing the data that has the potential to yield more information about the masses and lifetimes of the neutral scalars and/or SUSY particles involved. As we will see, there is a good possibility of measuring the neutralino mass using a simple technique that could be used in other experiments.

\section{Analysis}
\label{Analysis}

To estimate the sensitivity and ability to measure observables in the data, we use {\sc PYTHIA} 6.4~\cite{pythia2} to simulate the various neutral scalar boson production using production of single scalar boson $\mathrm{\varphi}$ with all combinations of the parameters listed above, and use a modified version of PGS4~\cite{PGS4} to simulate the CDF detector. Since any measurement is crucially dependent on the timing distribution, we implemented timing response as a modification to PGS4, using the prescriptions of Ref.~\cite{Mason:2011zv,PhysRevLett.99.121801,*PhysRevD.78.032015,SearchProspect,cdf10788.2,*JonThesis,*AdamThesis,Goncharov2006543}, which are known to be very Gaussian distribution responsive, even at large times. We use the same selection requirements as in the reported CDF result when simulating the detector~\cite{cdf10788.2,*JonThesis,*AdamThesis} as it allows for acceptance and background calculations.

\begin{figure}[htbp]
\begin{tabular}{c}
\includegraphics[width=0.5\textwidth]{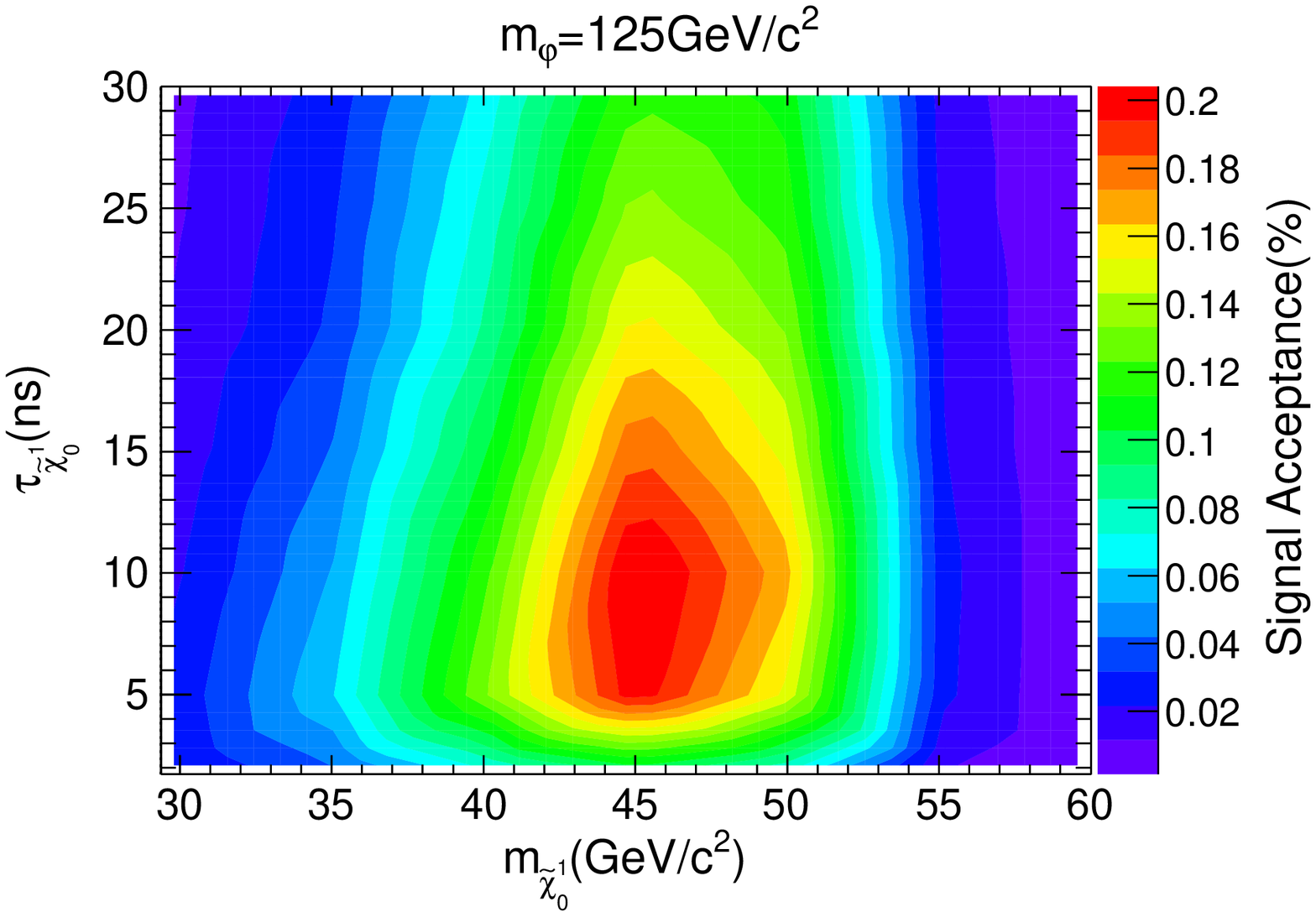}
\end{tabular}
\caption{The signal acceptance for $\mathrm{\varphi} \rightarrow \widetilde{\chi}^{1}_{0} \widetilde{\chi}^{1}_{0} \rightarrow \gamma_{Delayed}+\slashed{E}_{T}$, with $m_{\mathrm{\varphi}} = 125~{\rm GeV/c^2}$. The signal acceptance peaks at $m_{\widetilde{\chi}^{1}_{0}} = 45 ~{\rm GeV/c^2}$ and $\tau_{\widetilde{\chi}^{1}_{0}} = 10$ ns, with the peak acceptance equal to $0.21$\%.}
\label{AccCont} 	
\end{figure}

Assuming $m_{\mathrm{\varphi}} = 125~{\rm GeV/c^2}$, the contour of the acceptance in the signal region vs. $m_{\widetilde{\chi}^{1}_{0}}$ and $\tau_{\widetilde{\chi}^{1}_{0}}$ is shown in \fig{AccCont}. The acceptance peaks at $m_{\widetilde{\chi}^{1}_{0}} = 45 ~{\rm GeV/c^2}$ and $\tau_{\widetilde{\chi}^{1}_{0}} = 10$~ns, which we will take as our benchmark point; the signal acceptance after each analysis selection requirement is shown in Table~\ref{CutAcc}. The peak signal acceptance is $0.21$\% and quickly falls as a function of $m_{\widetilde{\chi}^{1}_{0}}$ as we get away from the peak which is slightly below $m_{\mathrm{\varphi}} > 2 \cdot m_{\widetilde{\chi}^{1}_{0}}$ because of the balance of the kinematics making the events pass the thresholds and the boost changing the photon delay. Similarly, as the lifetime rises, more and more of the neutralinos leave the detector so we lose sensitivity  above $\tau_{\widetilde{\chi}^{1}_{0}} >$  30 ns. As the lifetime goes below a few ns, fewer and fewer of the events have a long enough lifetime to produce a delayed photon that shows up in the signal region.

\begin{table}
\centering
\begin{tabular}{|c|c|} \hline
 Selection Requirement & Signal Acceptance  \\ \hline
$|\eta^{\gamma}|<1.1$ &  15.7\% \\ \hline
 $E_{T}^{\gamma}>45~{\rm GeV}$ & 2.58\% \\ \hline
 $\slashed{E}_{T}>45~{\rm GeV}$ & 1.53\% \\ \hline
 Jet and Track Veto  & 1.15\% \\ \hline
 $2~{\rm ns}<t_{corr}<7~{\rm ns}$ & 0.21\% \\ \hline
\end{tabular}
\caption{ The percentage of events passing each acceptance requirement (signal acceptance) for our benchmark parameter point of $m_{\mathrm{\varphi}} = 125~{\rm GeV/c^2}$, $m_{\widetilde{\chi}^{1}_{0}} = 45 ~{\rm GeV/c^2}$ and $\tau_{\widetilde{\chi}^{1}_{0}} = 10$ ns. We take into account a 75\% efficiency for the jet and track veto from~\cite{Mason:2011zv} since it is not well modeled in PGS.}
\label{CutAcc}
\end{table}

If we calculate the number of events we expect from our model, assuming $10~{\rm fb^{-1}}$ data, and taking our assumption of $\sigma_{\widetilde{\chi}^{1}_{0} \widetilde{\chi}^{1}_{0}}$ $\sim$0.5~pb, we end up with $\sim$10 events in the signal region at the peak acceptance.

\begin{figure}[htbp]
\begin{tabular}{c}
\includegraphics[width=0.5\textwidth]{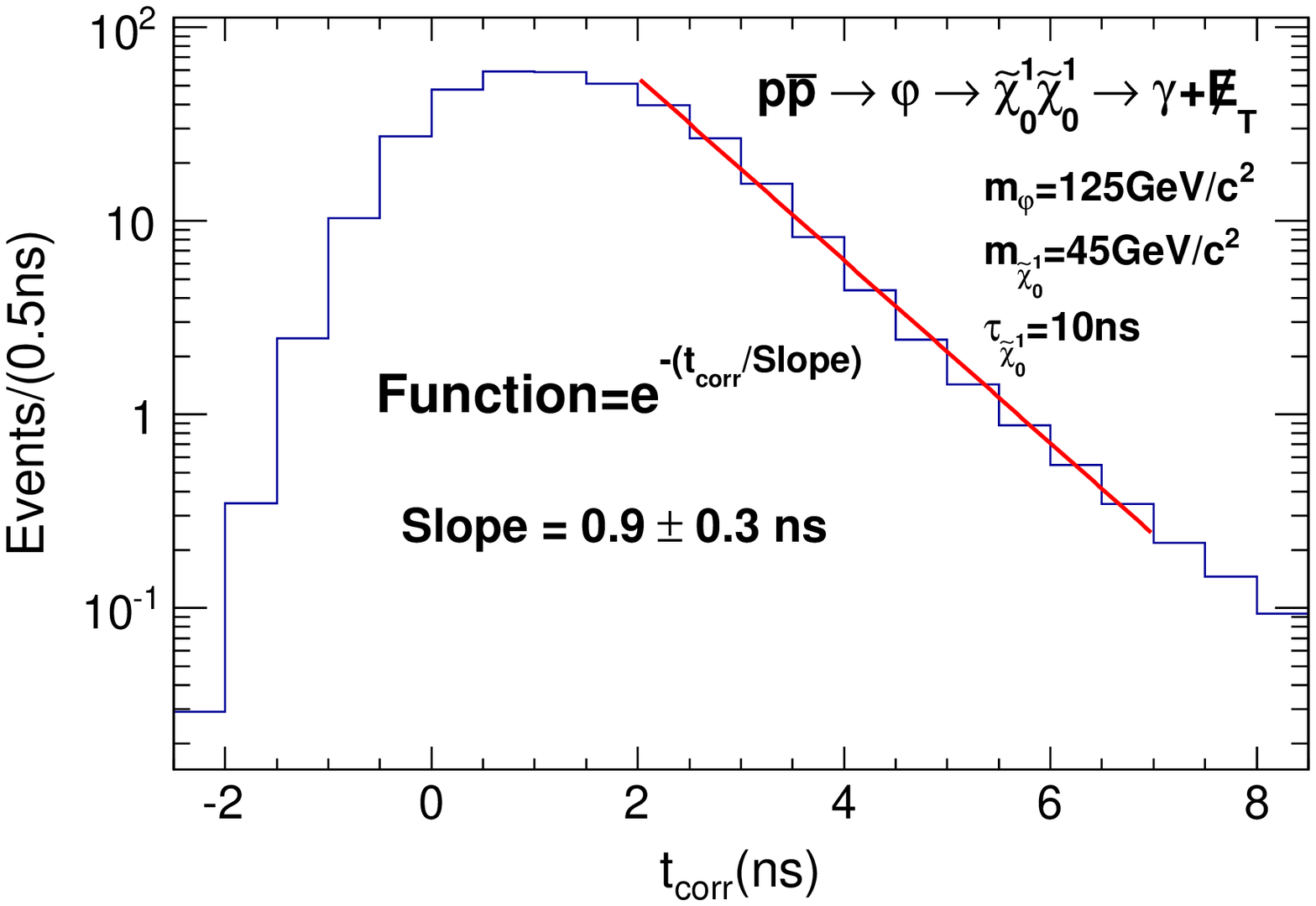}
\end{tabular}
\caption{The $t_{corr}$ distribution for our benchmark point, with $m_{\mathrm{\varphi}} = 125~{\rm GeV/c^2}$, $m_{\widetilde{\chi}^{1}_{0}} = 45 ~{\rm GeV/c^2}$ and $\tau_{\widetilde{\chi}^{1}_{0}} = 10$ ns. We assume $\sigma_{\widetilde{\chi}^{1}_{0} \widetilde{\chi}^{1}_{0}}$ = 0.5~pb and normalize the histogram so that there are 10 events in the signal region. Note that within the signal region the signal is well described by an exponential distribution.}
\label{tcorr} 	
\end{figure}

The full timing distribution of the signal can be quite complicated by the geometry of the detector. However, we have found that within the signal region chosen, to an excellent degree of approximation, the shape of the signal is well described by an exponential function in the signal region, as shown in \fig{tcorr}. Remarkably, this is true for the entire parameter space we consider in this timing window. This allows for a more sophisticated analysis other than just the counting experiment which is dominated by the number of events with $t_{corr}$ just above 2~ns. We can readily fit the signal region to the functional form $e^{-t_{corr}/Slope}$. Though the CDF result doesn't report a slope, we can see what the theory predicts. For our benchmark point we find a slope of $\sim$0.9 ns. With our expectation of approximately 10 events in the signal region, the fit would give us an approximate statistical uncertainty on the slope of $\sim$30\%.

\begin{figure}[htbp]
\begin{tabular}{c}
\includegraphics[width=0.5\textwidth]{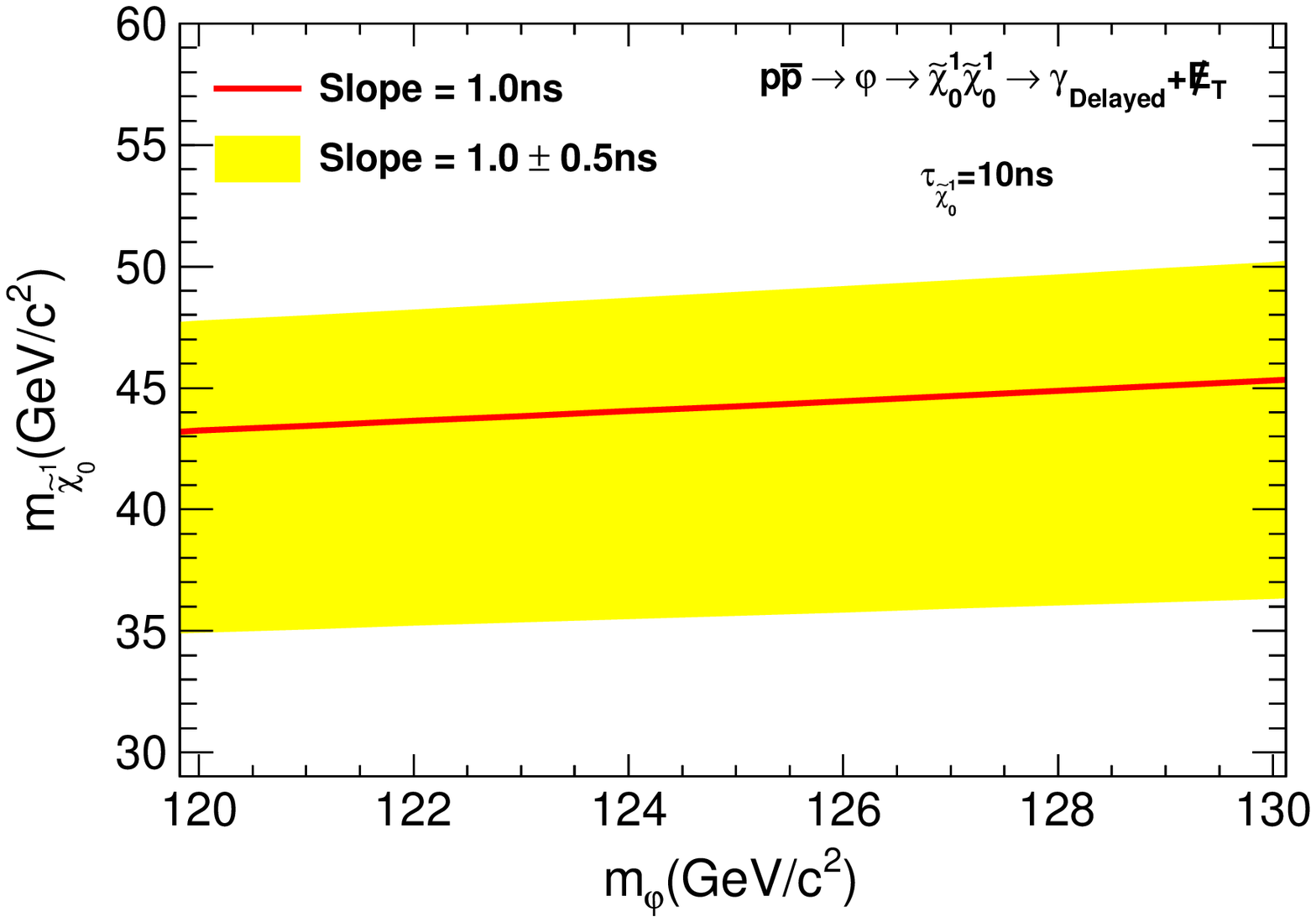}
\end{tabular}
\caption{Assuming $\tau_{\widetilde{\chi}^{1}_{0}} = 10$ ns, the red line shows possible combinations of $m_{\mathrm{\varphi}}$ and $m_{\widetilde{\chi}^{1}_{0}}$ that produce a slope of 1.0~ns, while the yellow band shows combinations that produce a slope of $1.0 \pm 0.5$ ns.}
\label{tau10} 	
\end{figure}

Luckily, simulations show that within the signal timing window, the exponential shape is maintained and the slope is a smooth function of $m_{\mathrm{\varphi}}$, $m_{\widetilde{\chi}^{1}_{0}}$, and $\tau_{\widetilde{\chi}^{1}_{0}}$ allowing for the exciting possibility of a clean measurement of these parameters. To get a sense of what parameters can be determined from a measurement we consider how the slope changes as a function of our three parameters as it does not depend on $\sigma_{\widetilde{\chi}^{1}_{0} \widetilde{\chi}^{1}_{0}}$. For concreteness we assume a measured slope of $1.0 \pm 0.5$~ns to both simplify the analysis as well as overestimate the uncertainty as a way of including systematic errors and see how well we can measure back the assumed input parameters that yield this combination, within uncertainties. Assuming $\tau_{\widetilde{\chi}^{1}_{0}} = 10$ ns, the possible combinations of $m_{\mathrm{\varphi}}$ and $m_{\widetilde{\chi}^{1}_{0}}$ is shown as the yellow band in \fig{tau10}. We quickly note that there is only a small variation as a function of $m_{\mathrm{\varphi}}$, which gives no sensitivity to measure the neutral scalar mass. We also note that if multiple scalars were to exist and contribute similarly to the signal region, we would further have no sensitivity to measure their masses or distinguish between the single or multiple scalar scenarios. The upside to this result is that since there is only modest variation as a function of the effective scalar mass, it validates our simplification of simulating with a single scalar, and gives us better sensitivity to the other two parameters.

\begin{figure}[htbp]
\begin{tabular}{c}
\includegraphics[width=0.5\textwidth]{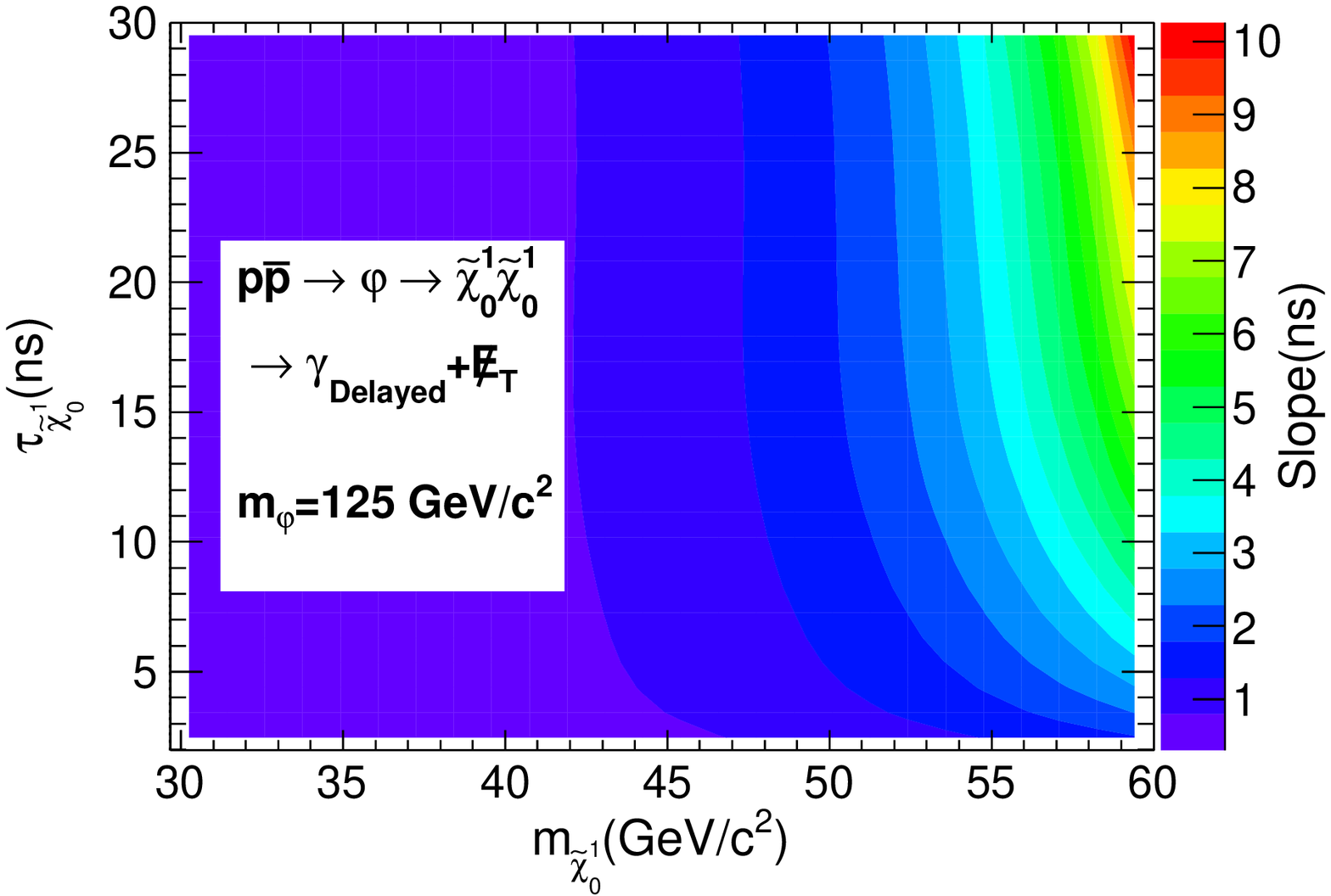}
\end{tabular}
\caption{Assuming $m_{\mathrm{\varphi}} = 125~{\rm GeV/c^2}$, this plot shows the contour of constant Slope as a function of $m_{\widetilde{\chi}^{1}_{0}}$ and $\tau_{\widetilde{\chi}^{1}_{0}}$. }
\label{slopecont} 	
\end{figure}

\begin{figure}[htbp]
\centering
\mbox{
\subfigure[]{
\includegraphics[width=0.46\textwidth]{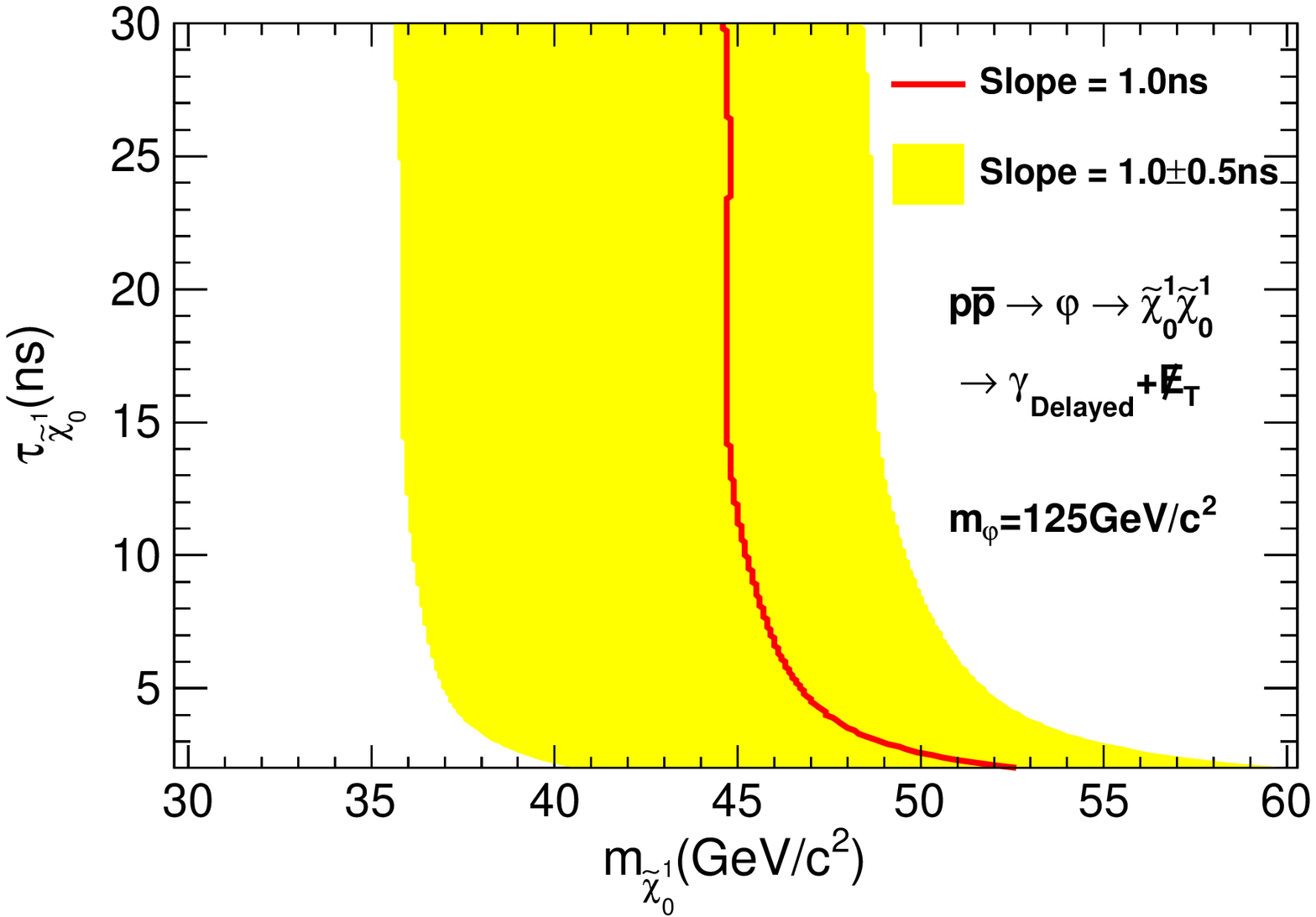}
\label{h125:h125sub1}}
\quad
\subfigure[]{
\includegraphics[width=0.46\textwidth]{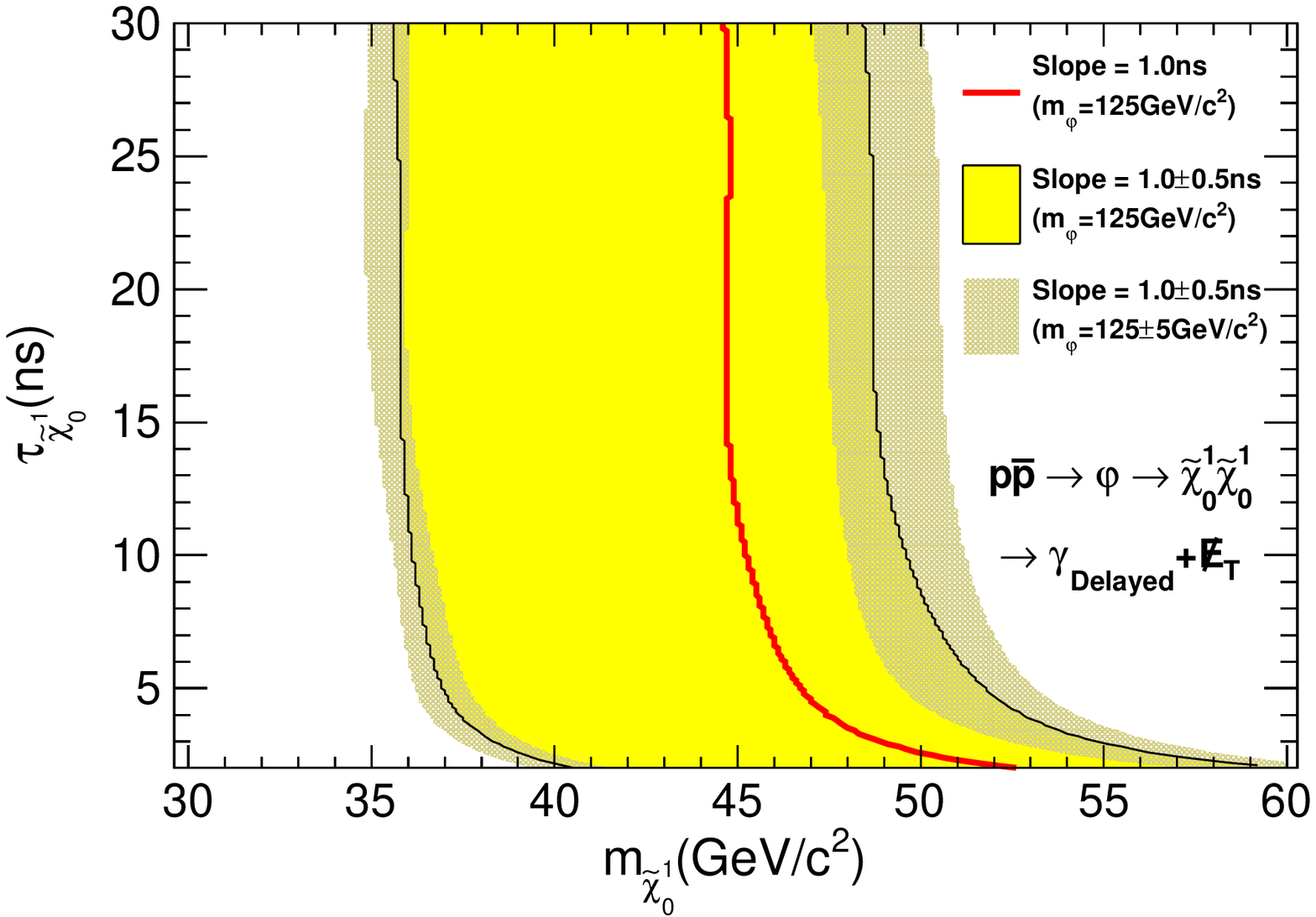}
\label{h125:h125sub2}}
}
\centering
\caption{Assuming $m_{\mathrm{\varphi}} = 125~{\rm GeV/c^2}$, the red line shows possible combinations of $m_{\widetilde{\chi}^{1}_{0}}$ and $\tau_{\widetilde{\chi}^{1}_{0}}$ that produce a slope of 1.0 ns, while the yellow band shows the combinations that produce a slope of $1.0 \pm 0.5$ ns. The grey band in (b) shows the variation due to the uncertainty of the effective scalar mass.}
\label{h125} 	
\end{figure}

Assuming $m_{\mathrm{\varphi}} = 125~{\rm GeV/c^2}$, \fig{slopecont} shows the magnitude of the slope as a function of $m_{\widetilde{\chi}^{1}_{0}}$ and $\tau_{\widetilde{\chi}^{1}_{0}}$. Note that the slope shows very little variation for lifetimes above about 5~ns. If we again assume that we had measured a slope of $1.0 \pm 0.5$~ns, taking into account an assumption of large systematic uncertainties, we narrow down the $m_{\widetilde{\chi}^{1}_{0}}$ and $\tau_{\widetilde{\chi}^{1}_{0}}$ combination to the yellow band shown in \fig{h125:h125sub1}. A $5~{\rm GeV/c^2}$ uncertainty of the effective scalar mass is shown by the grey band in \fig{h125:h125sub2}.

\begin{figure}[htbp]
\begin{tabular}{c}
\includegraphics[width=0.5\textwidth]{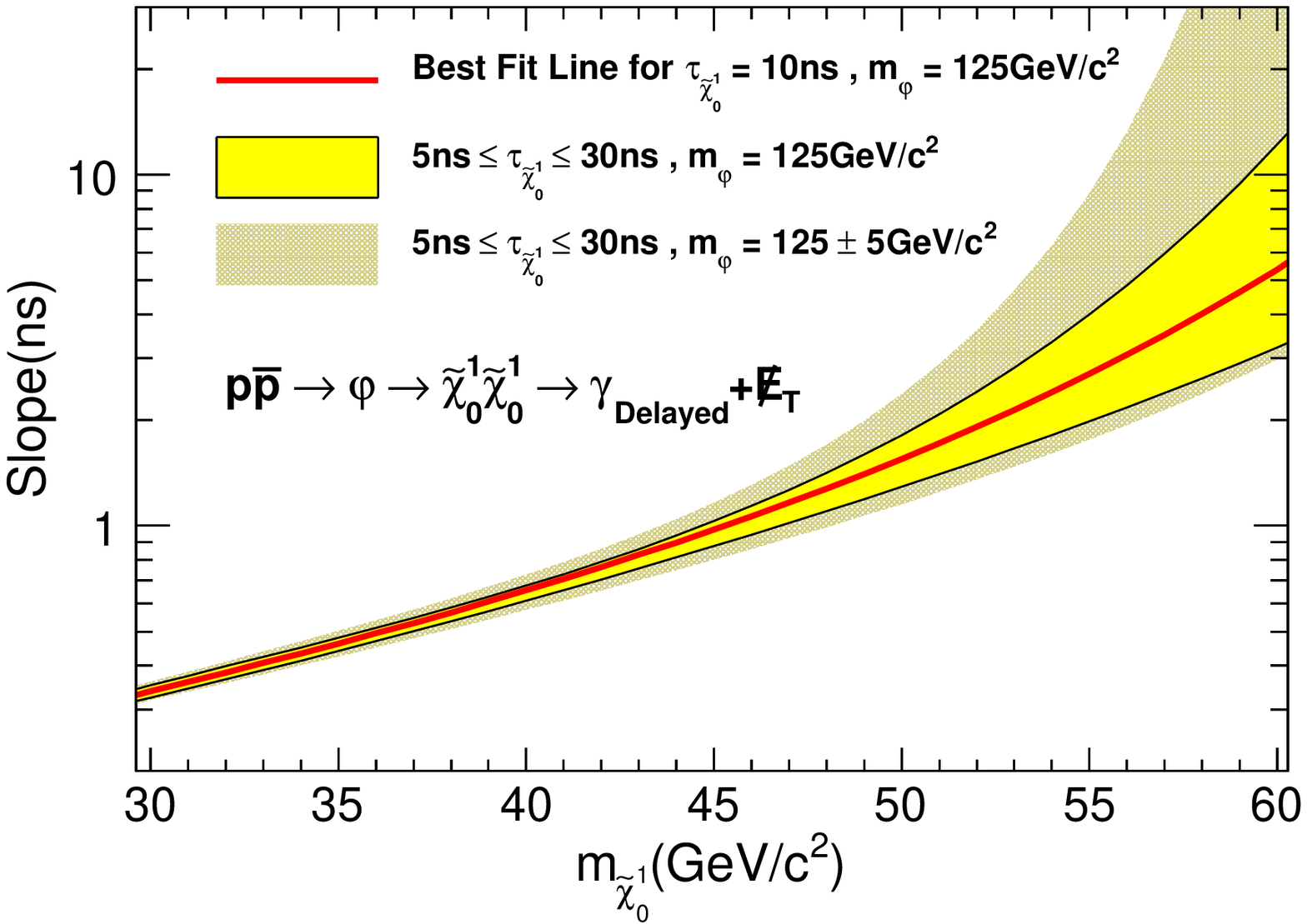}
\end{tabular}
\caption{Assuming $\tau_{\widetilde{\chi}^{1}_{0}} \ge 5$ns, the simple dependence between the slope and $m_{\widetilde{\chi}^{1}_{0}}$.}
\label{schi} 	
\end{figure}

Since there is no evidence for short neutralino lifetime in the $\gamma\gamma+\slashed{E}_{T}$ search~\cite{PhysRevD.82.052005}, if we further assume $\tau_{\widetilde{\chi}^{1}_{0}} \ge 5$ns, we observe a simple dependence between the slope and the neutralino mass, as shown in \fig{schi}. If we had measured a slope of $1.0 \pm 0.5$ ns, from here we could determine the neutralino mass to be $m_{\widetilde{\chi}^{1}_{0}} = 45^{+8}_{-10}~{\rm GeV/c^2}$. Of all the possibilities, we have found that our simple model of $\mathrm{\varphi} \rightarrow \widetilde{\chi}^{1}_{0} \widetilde{\chi}^{1}_{0} \rightarrow (\gamma \widetilde{G})(\gamma \widetilde{G}) \rightarrow \gamma_{Delayed}+\slashed{E}_{T}$ has a reconstructed timing distribution that is simple enough to be measured in an experiment, and that despite the potential for complications between multiple parameters it allows for straightforward measurement of the neutralino mass. Said differently, our measurement of the slope of the timing distribution allows for a novel method to measure $m_{\widetilde{\chi}^{1}_{0}}$ despite how complicated it could have been.

\section{Conclusion}
\label{Conclusion}
We have described a method to measure the mass of a heavy, neutral particle with $\mathcal{O}$(10~ns) lifetime, which decays to photons in a collider experiment, using the time of arrival distribution of a sample of delayed photons, and estimated the potential sensitivity for any measurement. In principle this method can be applied to any model with a signature of a delayed photon in its final state. In this study we focused on a LNG-GMSB SUSY scenario where sparticle production proceeds through neutral scalar bosons at CDF as we have a full set of tools that allow for a reliable prospects study. Following the selection requirements of the CDF preliminary results, we find a peak signal acceptance of 0.21\% at $m_{\mathrm{\varphi}} = 125~{\rm GeV/c^2}$, $m_{\widetilde{\chi}^{1}_{0}} = 45 ~{\rm GeV/c^2}$, and $\tau_{\widetilde{\chi}^{1}_{0}} = 10$~ns. With a 0.5~pb production cross section, this would give roughly 10 events and would yield a measurement of the slope of the $t_{corr}$ distribution in the signal region. In this case, we would have sensitivity to measure the mass of neutralino with modest assumptions. Even a 50\% measurement of the slope could give a measurement of the neutralino mass with an uncertainty of ~25\%, depending on the true parameters involved.

We encourage the CDF collaboration to update their search in the exclusive $\gamma_{Delayed}+\slashed{E}_{T}$ final state using the full dataset and to report a measurement for the slope of the timing distribution in the case that the excess holds up. In addition, we encourage other experiments to confirm or refute these preliminary observations and present their background and signal estimation methods in a manner that allows for sensitivity studies. Our preliminary studies of the phenomenology for LHC experiments indicate that while the production cross sections would clearly be higher, beam-spot size effects would make wrong-vertex backgrounds smaller, and the timing from the detectors could produce better sensitivity, the pileup effects and the larger boosts of the $\widetilde{\chi}^{1}_{0}$ could cancel out many of these advantages. However, since the possibilities are always exciting, we encourage them to provide background estimates and tools to estimate the sensitivity as soon as possible so we can perform further sensitivity studies to see if our promising results can be used there as well.

\section{Acknowledgements}

The authors would like to thank Adam Aurisano, Bhaskar Dutta and John Mason for their useful ideas and discussions. We would also thank the Mitchell Institute for Fundamental Physics and Astronomy and the Department of Physics and Astronomy at Texas A\&M University for their support during the time this work was completed. We acknowledge the Texas A\&M University Brazos HPC cluster that contributed to the research reported here\footnote{brazos.tamu.edu}.


\bibliographystyle{apsrev4-1}

\bibliography{PhotonMetCitations}

\end{document}